\documentclass[showkeys,superscriptaddress,floatfix,prd,10pt,aps]{revtex4-2}
\usepackage{graphicx,epstopdf}
\usepackage[caption=false]{subfig}
\usepackage{appendix}
\usepackage[T1]{fontenc}
\usepackage{lmodern}
\usepackage[dvipsnames,x11names]{xcolor}
\usepackage[colorlinks=true,linkcolor=NavyBlue,citecolor=ForestGreen,urlcolor=NavyBlue]{hyperref}
\usepackage[sort&compress]{natbib}
\usepackage{lipsum}
\usepackage{morefloats}
\usepackage[pdf]{pstricks}
\usepackage{amsmath,amsthm}
\usepackage{amssymb}
\usepackage{amsfonts}
\usepackage{rotating}
\usepackage{cancel}
\usepackage{mathtools}
\usepackage{bbm}
\usepackage{dsfont}
\usepackage{bbold}
\usepackage{multirow}
\usepackage{ulem}
\usepackage{physics}
\usepackage{orcidlink}
\usepackage{colortbl}

\def\pslash{p\!\!\!\slash }
\def\qslash{q\!\!\!\slash }
\def\xslash{x\!\!\!\slash }
\def\eslash{\varepsilon\!\!\!\slash }

\def\vel{\left|}
\def\ver{\right|}

\begin{document}

\title{Investigation on the electromagnetic properties of the $ D^{(*)} \Sigma_c^{(*)}$ molecules}

\author{Ula\c{s} \"{O}zdem\orcidlink{0000-0002-1907-2894}}%
\email[]{ulasozdem@aydin.edu.tr}
\affiliation{Health Services Vocational School of Higher Education, Istanbul Aydin University, Sefakoy-Kucukcekmece, 34295 Istanbul, T\"{u}rkiye}

 
\begin{abstract}
We systematically explore their electromagnetic characteristics to improve our understanding of the quark-gluon dynamics underlying the complex and controversial nature of multiquark systems. In this study, the magnetic dipole moments of $ D \Sigma_c$, $ D \Sigma_c^{*}$ and $ D^{*} \Sigma_c$ doubly-charmed pentaquarks are extracted, which are directly related to the inner organization of the relevant states. The magnetic dipole moments of these states have been evaluated employing the QCD light-cone sum rules technique with isospin spin-parity $\rm{I(J^P)} = \frac{1}{2}(\frac{1}{2}^-)$, $\rm{I(J^P)} = \frac{1}{2}(\frac{3}{2}^-)$ and $\rm{I(J^P)} = \frac{1}{2}(\frac{3}{2}^-)$,  for $ D \Sigma_c$, $ D \Sigma_c^{*}$ and $ D^{*} \Sigma_c$ doubly-charmed pentaquarks respectively. Our predictions for the magnetic dipole moment  $\mu_{D\Sigma_c} = 2.98^{+0.76}_{-0.54}~\mu_N$ for the $ D \Sigma_c$ pentaquark,  $\mu_{D\Sigma_c^*} = 1.65^{+0.45}_{-0.34}~\mu_N$ for the $D\Sigma_c^*$ pentaquark, and $\mu_{D^*\Sigma_c} = -3.63^{+0.79}_{-0.60}~\mu_N$  for the $D^*\Sigma_c$ pentaquark. Furthermore, we have also extracted the electric quadrupole and the magnetic octupole moments of the $ D \Sigma_c^{*}$ and $ D^{*} \Sigma_c$ doubly-charmed pentaquarks. These values show a non-spherical charge distribution.  As a by-product, the magnetic dipole moments of the isospin-$\frac{3}{2}$ partners associated with these doubly-charmed pentaquarks have also been determined. The magnetic dipole moments are calculated as follows: $\mu_{D\Sigma_c} = 3.78^{+0.94}_{-0.70}~\mu_N$, $  \mu_{D\Sigma_c^*} = 2.08^{+0.57}_{-0.43}~\mu_N$,  $\mu_{D^*\Sigma_c} = -4.59^{+0.99}_{-0.76}~\mu_N$ for the isospin-$\frac{3}{2}$ partners $ D \Sigma_c$, $ D \Sigma_c^{*}$ and $ D^{*} \Sigma_c$ doubly-charmed pentaquarks, respectively. We hope that our predictions of the magnetic dipole moments of the doubly-charmed pentaquarks, in conjunction with the results of other theoretical investigations of the spectroscopic parameters and decay widths of these intriguing pentaquarks, will prove valuable in the search for these states in future experiments and in elucidating the internal structure of these pentaquarks. 
\end{abstract}

\maketitle

\section{Introduction}\label{motivation}

The conventional quark model is unable to explain the exotic hadrons, such as the compact multiquark states and hadronic molecular states. These states have become a central topic in hadron physics and have been the subject of considerable research over the past three decades. In particular, several hidden-charm pentaquarks have been discovered, including the $\mathrm{P_{c}(4380)}$, $\mathrm{P_{c}(4450)}$, $\mathrm{P_{c}(4440)}$, $\mathrm{P_{c}(4457)}$, $\mathrm{P_{c}(4312)}$, $\mathrm{P_{cs}(4459)}$, and $\mathrm{P_{cs}(4338)}$~\cite{Aaij:2015tga,Aaij:2019vzc,LHCb:2021chn,Collaboration:2022boa}. Given that these states contain at least five quarks ($c\bar c udd$ or $c\bar c uds$), they are ideal candidates for the hidden-charm pentaquarks. In 2021, the LHCb Collaboration reported observation of a first doubly charmed tetraquark state $ T_{cc}^+(3875)$  in the $D^0 D^0 \pi^+$ mass spectrum~\cite{LHCb:2021vvq,LHCb:2021auc}.  The simplest assumption on its valence quark component is $c c\bar u \bar d$ and spin-parity quantum numbers were determined to be $\rm{J^P} = 1^+$. Following the experimental discoveries, several theoretical interpretations were put forth to elucidate them, including the compact states, hadronic molecular states, kinematical effects, and so on~\cite{Esposito:2014rxa,Esposito:2016noz,Olsen:2017bmm,Lebed:2016hpi,Nielsen:2009uh,Brambilla:2019esw,Agaev:2020zad,Chen:2016qju,Ali:2017jda,Guo:2017jvc,Liu:2019zoy,Yang:2020atz,Dong:2021juy,Meng:2022ozq, Chen:2022asf,Dong:2021bvy}. Nevertheless, despite the considerable efforts made, the substructures and quantum numbers of these multiquark states remain one of the most challenging and significant questions in high-energy physics.

The discovery of the $ T_{cc}^+(3875)$ tetraquark state will provide a novel avenue for the search for additional states beyond the conventional hadrons. If the $ T_{cc}^+(3875)$ is indeed the doubly-charmed tetraquark state, it is also possible that the doubly-charmed pentaquarks may exist. The aforementioned argument follows a similar logic to that which was employed in the prediction of the hidden-charm pentaquarks, given the existence of the hidden-charm tetraquarks.  In light of the lack of evidence supporting the existence of doubly-charmed pentaquarks, it becomes crucial to investigate the reasons behind this apparent absence.  To date, theoretical studies have been conducted to examine the potential existence of doubly-charmed pentaquarks~\cite{Chen:2017vai,Liu:2020nil,Chen:2021kad,Chen:2021htr,Dong:2021bvy,Shen:2022zvd,Liu:2023clr,Yang:2020twg,Shimizu:2017xrg,Guo:2017vcf,Zhou:2018bkn,Wang:2018lhz,Xing:2021yid,Duan:2024uuf,Wang:2024brl,Wang:2022ltr,Wang:2022gfb}. The results of the studies conducted indicate that the doubly-charmed pentaquarks are lie below the relevant meson-baryon threshold and they maybe form the bound states. In comparison to doubly-charm baryons, doubly-charm pentaquarks are expected to have a larger mass.  Nevertheless, the intricate interactions within multiquark states may result in a reduction in mass, potentially complicating the distinction between traditional baryons and pentaquarks solely based on mass. An investigation of these states' other properties, including electromagnetic properties, may provide insights into their internal composition.  

 The electromagnetic multipole moments represent measurable quantities that may offer significant insights into the quark-gluon composition of hadrons and their fundamental nature, including their underlying dynamics. In the electro- or photo-production of pentaquarks, the magnetic dipole moments of the pentaquarks have a significant impact on both the differential and total cross-sections. Consequently, it is of paramount importance to determine the magnetic dipole moments of the pentaquarks to verify their nature. While the electromagnetic properties of pentaquarks have not been as extensively studied experimentally or theoretically as their mass spectra, production properties, quantum numbers, and decay behavior, they have gained increasing interest as key parameters in understanding states with controversial internal structures~\cite{Wang:2016dzu, Ozdem:2018qeh, Ortiz-Pacheco:2018ccl,Xu:2020flp,Ozdem:2021btf, Li:2021ryu,Ozdem:2021ugy,Gao:2021hmv, Ozdem:2022iqk, Ozdem:2022vip, Ozdem:2022kei, Wang:2022tib,Ozdem:2023htj,Wang:2023iox,Wang:2023ael,Guo:2023fih,Ozdem:2024rqx,Ozdem:2024jty,Lai:2024jfe,Huang:2024iua,Li:2024wxr,Mutuk:2024jxf,Zhou:2022gra}. In this study, therefore,  we employ the QCD light-cone sum rules to study the magnetic dipole moments of the $ D^{(*)} \Sigma_c^{(*)}$ states, considering that these pentaquarks are $ D \Sigma_c$, $ D^{*} \Sigma_c$, and $ D \Sigma_c^{*}$ molecular configurations with isospin-spin-parity $\mathrm{I(J^P)= \frac{1}{2}(\frac{1}{2}^-)}$, $\mathrm{I(J^P)= \frac{1}{2}(\frac{3}{2}^-})$, and $\mathrm{I(J^P)= \frac{1}{2}(\frac{3}{2}^-}$), respectively. Among the numerous tools available for calculating non-perturbative parameters, the QCD light-cone sum rule represents a highly efficient approach~\cite{Chernyak:1990ag, Braun:1988qv, Balitsky:1989ry}. In the context of QCD light-cone sum rules, the correlation function will be calculated on two sides. The first of these is referred to as the "hadronic side", while the second is the "QCD side". According to the principle of quark-hadron duality, computations of these two representations will be coordinated. This enables the derivation of sum rules for magnetic dipole moments at the hadronic level.

This paper is organized as follows. After the introduction, the QCD light-cone sum rules for the magnetic dipole moments of the $ D \Sigma_c$, $ D^{*} \Sigma_c$, and $ D \Sigma_c^{*}$ pentaquarks are derived in Sec. \ref{formalism}. In Sec. \ref{numerical}, we present the corresponding numerical results. The manuscript ends with a summary in Sec. \ref{summary}. 

\begin{widetext}
 
\section{The method of calculating the magnetic dipole moments}\label{formalism}

The QCD light-cone sum rules for the magnetic dipole of 
the spin-1/2 and spin-3/2 doubly-charmed pentaquarks ($\mathrm{P_{cc}} \rightarrow \rm{J^P}=\frac{1}{2}^-$, $\mathrm{P_{cc}^*}\rightarrow \rm{J^P}=\frac{3}{2}^-$) can be extracted from the analysis of the following two-point correlation functions in a weak external electromagnetic field, denoted by $\gamma$,
 \begin{align} \label{edmn01}
\Pi(p,q)&=i\int d^4x e^{ip \cdot x} \langle0|T\left\{\rm{J}^{\mathrm{P_{cc}}}(x)\bar{\rm{J}}^{\mathrm{P_{cc}}}(0)\right\}|0\rangle _\gamma \, , \\
\Pi_{\mu\nu}(p,q)&=i\int d^4x e^{ip \cdot x} \langle0|T\left\{\rm{J}_\mu^{\mathrm{P_{cc}^*}}(x)\bar{\rm{J}}_\nu^{\mathrm{P_{cc}^*}}(0)\right\}|0\rangle _\gamma \,, \label{Pc101}
\end{align}
where ${\mathrm{J^{P_{cc}}}}(x)$ and $\mathrm{J_\mu^{P^*_{cc}}}(x)$ are the interpolating currents of the corresponding states. 
The relevant interpolating currents required for our analysis in the framework of the QCD light cone sum rule are given by the formulas 
\begin{align}
\label{curpcs2}
\rm{J}^{ D \Sigma_c}(x)& =\frac{1}{\sqrt{3}}\mid  D^+(x) \Sigma_c^+(x) \rangle \, \mp \sqrt{\frac{2}{3}}\mid D^0(x) \Sigma_c^{++}(x) \rangle  
= \frac{1}{\sqrt{3}}  \big[- \bar d^d(x)i \gamma_5 c^d(x)\big]\big[\varepsilon^{abc} u^{a^T}(x)C\gamma_\alpha d^b(x)  \gamma^\alpha\gamma_5 c^c(x)\big]\nonumber\\
&
 \mp \sqrt{\frac{2}{3}} \big[\bar u^d(x)i \gamma_5 c^d(x)\big]  
 \big[\varepsilon^{abc} u^{a^T}(x) C\gamma_\alpha u^b(x) 
 \gamma^\alpha\gamma_5 c^c(x)\big] \, , \\
\rm{J}_\mu^{ D \Sigma_c^*}(x)&=\frac{1}{\sqrt{3}}\mid  D^+(x) \Sigma_c^{*+}(x) \rangle \, \mp \sqrt{\frac{2}{3}}\mid  D^0(x) \Sigma_c^{*++}(x) \rangle  
= \frac{1}{\sqrt{3}}  \big[-\bar d^d(x)i \gamma_5 c^d(x)\big]\big[\varepsilon^{abc} u^{a^T}(x)C\gamma_\mu d^b(x)  c^c(x)\big]\nonumber\\
&
 \mp \sqrt{\frac{2}{3}} \big[\bar u^d(x)i \gamma_5 c^d(x)\big]  
 \big[\varepsilon^{abc} u^{a^T}(x) C\gamma_\mu u^b(x) 
  c^c(x)\big] \, , \\  
\rm{J}_\mu^{ D^* \Sigma_c}(x)& =\frac{1}{\sqrt{3}}\mid  D^{*+}(x) \Sigma_c^+(x) \rangle \, \mp \sqrt{\frac{2}{3}}\mid  D^{*0}(x) \Sigma_c^{++}(x) \rangle  
= \frac{1}{\sqrt{3}}  \big[-\bar d^d(x)\gamma_\mu c^d(x)\big]\big[\varepsilon^{abc} u^{a^T}(x)C\gamma_\alpha d^b(x)  \gamma^\alpha\gamma_5 c^c(x)\big]\nonumber\\
&
 \mp \sqrt{\frac{2}{3}} \big[\bar u^d(x)\gamma_\mu c^d(x)\big]  
 \big[\varepsilon^{abc} u^{a^T}(x) C\gamma_\alpha u^b(x) 
 \gamma^\alpha\gamma_5 c^c(x)\big]\,. 
  \end{align}
   
  As can be observed, the "$\mp$" symbol is represented in the explicit form of the interpolating currents presented above. If the sign between these currents is negative, it couples to the isospin-1/2 states, whereas if it is positive, it couples to the isospin-3/2 states.  In this study, the corresponding interpolating currents are constructed with the definite isospins to couple with the color singlet-singlet type doubly-charmed pentaquarks rather than with undesired effects such as the meson-baryon scattering states or thresholds. This crucial step in resolving the puzzle surrounding the doubly-charmed pentaquarks serves to enhance the precision and reliability of these analyses. 

Following the methodology of the QCD light-cone sum rule, the correlation function is first expressed as a function of the physical parameters of the hadron. In the context of hadron parameters, a complete set of hadronic states with the same quantum numbers as the interpolating currents $\mathrm{J^{P_{cc}}}(x)$ and $\mathrm{J_\mu^{P^*_{cc}}}(x)$ are inserted into the correlation functions. The dispersion relation allows for the correlation functions to be written as
 \begin{align}\label{edmn02}
\Pi^{Had}(p,q)&=\frac{\langle0\mid \rm{J}^{\mathrm{P_{cc}}}(x) \mid
{\mathrm{P_{cc}}}(p, s) \rangle}{[p^{2}-m_{\mathrm{P_{cc}}}^{2}]}
\langle {\mathrm{P_{cc}}}(p, s)\mid
{\mathrm{P_{cc}}}(p+q, s)\rangle_\gamma 
\frac{\langle {\mathrm{P_{cc}}}(p+q, s)\mid
\bar{ \rm{J}}^{\mathrm{P_{cc}}}(0) \mid 0\rangle}{[(p+q)^{2}-m_{\mathrm{P_{cc}}}^{2}]}+ \cdots , \\
\nonumber\\ 
\Pi^{Had}_{\mu\nu}(p,q)&=\frac{\langle0\mid  \rm{J}_{\mu}^{\mathrm{P_{cc}^*}}(x)\mid
{\mathrm{P_{cc}^*}}(p,s)\rangle}{[p^{2}-m_{{\mathrm{P_{cc}^*}}}^{2}]}
\langle {\mathrm{P_{cc}^*}}(p,s)\mid
{\mathrm{P_{cc}^*}}(p+q,s)\rangle_\gamma 
\frac{\langle {\mathrm{P_{cc}^*}}(p+q,s)\mid
\bar{\rm{J}}_{\nu}^{\mathrm{P_{cc}^*}}(0)\mid 0\rangle}{[(p+q)^{2}-m_{{\mathrm{P_{cc}^*}}}^{2}]}+ \cdots .\label{Pc103}
\end{align}

To perform further calculations, it is necessary to have access to the matrix elements specified in Eqs.~(\ref{edmn02}) and (\ref{Pc103}). These elements can be expressed in terms of the hadronic parameters as follows:
%
\begin{align}
\label{edmn005}
\langle0\mid \rm{J}^{\mathrm{P_{cc}}}(x)\mid {\mathrm{P_{cc}}}(p, s)\rangle=&\lambda_{\mathrm{P_{cc}}} \gamma_5 \, \nu(p,s),\\
\langle {\mathrm{P_{cc}}}(p+q, s)\mid \rm{\bar J}^{\mathrm{P_{cc}}}(0)\mid 0\rangle=&\lambda_{\mathrm{P_{cc}}} \gamma_5 \, \bar \nu(p+q,s)\\
\langle {\mathrm{P_{cc}}}(p, s)\mid {\mathrm{P_{cc}}}(p+q, s)\rangle_\gamma &=\varepsilon^\mu\,\bar \nu(p, s)\bigg[\big[f_1(q^2)
+f_2(q^2)\big] \gamma_\mu +f_2(q^2)
\frac{(2p+q)_\mu}{2 m_{\mathrm{P_{cc}}}}\bigg]\,\nu(p+q, s), \\
\nonumber\\
\langle0\mid \rm{J}_{\mu}^{\mathrm{P_{cc}^*}}(x)\mid {\mathrm{P_{cc}^*}}(p,s)\rangle&=\lambda_{{\mathrm{P_{cc}^*}}}u_{\mu}(p,s),\\
\langle {\mathrm{P_{cc}^*}}(p+q,s)\mid   \rm{\bar{J}}_{\nu}^{\mathrm{P_{cc}^*}}(0)\mid 0\rangle &= \lambda_{{\mathrm{P_{cc}^*}}}\bar u_{\nu}(p+q,s), \\
\langle {\mathrm{P_{cc}^*}}(p,s)\mid {\mathrm{P_{cc}^*}}(p+q,s)\rangle_\gamma &=-e\bar u_{\mu}(p,s)\bigg[F_{1}(q^2)g_{\mu\nu}\eslash 
-
\frac{1}{2m_{{\mathrm{P_{cc}^*}}}} 
\Big[F_{2}(q^2)g_{\mu\nu} 
+F_{4}(q^2)\frac{q_{\mu}q_{\nu}}{(2m_{{\mathrm{P_{cc}^*}}})^2}\Big]\eslash\qslash
\nonumber\\
&+
F_{3}(q^2)\frac{1}{(2m_{{\mathrm{P_{cc}^*}}})^2}q_{\mu}q_{\nu}\eslash \bigg] 
u_{\nu}(p+q,s),
\label{matelpar}
\end{align}
where $ \nu(p+q,s)$  and $u_{\mu}(p,s)$ are the spinors of the $\mathrm{P_{cc}}$ state and $\mathrm{P_{cc}^*}$ state, respectively, and; $f_i$'s and $F_i$'s are the form factors of the related transitions.

With the help of the above equations, we obtain the following expressions for the hadronic representation of the correlation functions for magnetic dipole moments: 
\begin{align}
\label{edmn05}
\Pi^{Had}(p,q)=&\lambda^2_{\mathrm{P_{cc}}}\gamma_5 \frac{\big(\pslash+m_{\mathrm{P_{cc}}} \big)}{[p^{2}-m_{{\mathrm{P_{cc}}}}^{2}]}\varepsilon^\mu \bigg[\big[f_1(q^2) %
+f_2(q^2)\big] \gamma_\mu
+f_2(q^2)\, \frac{(2p+q)_\mu}{2 m_{\mathrm{P_{cc}}}}\bigg]  \gamma_5 
\frac{\big(\pslash+\qslash+m_{\mathrm{P_{cc}}}\big)}{[(p+q)^{2}-m_{{\mathrm{P_{cc}}}}^{2}]}, \\
\nonumber\\
\Pi^{Had}_{\mu\nu}(p,q)&=-\frac{\lambda_{_{\mathrm{P_{cc}^*}}}^{2}\,\big(\pslash+\qslash+m_{\mathrm{P_{cc}^*}}\big)}{[(p+q)^{2}-m_{_{\mathrm{P_{cc}^*}}}^{2}][p^{2}-m_{_{\mathrm{P_{cc}^*}}}^{2}]}
 \bigg[g_{\mu\nu}
-\frac{1}{3}\gamma_{\mu}\gamma_{\nu}-\frac{2\,p_{1\mu}p_{1\nu}}
{3\,m^{2}_{\mathrm{P_{cc}^*}}}+\frac{(p+q)_{\mu}\gamma_{\nu}-(p+q)_{\nu}\gamma_{\mu}}{3\,m_{\mathrm{P_{cc}^*}}}\bigg]   \nonumber\\
& \times  \bigg\{F_{1}(q^2)g_{\mu\nu}\eslash - 
\frac{1}{2m_{\mathrm{P_{cc}^*}}}
\Big[F_{2}(q^2)g_{\mu\nu} +F_{4}(q^2) \frac{q_{\mu}q_{\nu}}{(2m_{\mathrm{P_{cc}^*}})^2}\Big]\eslash\qslash+\frac{F_{3}(q^2)}{(2m_{\mathrm{P_{cc}^*}})^2}
 q_{\mu}q_{\nu}\eslash\bigg\}
 \nonumber\\
 &\times
 \big(\pslash+m_{\mathrm{P_{cc}^*}}\big)
 \bigg[g_{\mu\nu}-\frac{1}{3}\gamma_{\mu}\gamma_{\nu}-\frac{2\,p_{\mu}p_{\nu}}
{3\,m^{2}_{\mathrm{P_{cc}^*}}}+\frac{p_{\mu}\gamma_{\nu}-p_{\nu}\gamma_{\mu}}{3\,m_{\mathrm{P_{cc}^*}}}\bigg].
\label{final phenpart}
\end{align}

The aforementioned equations for the hadronic representation can be expressed in a more concise and straightforward manner as follows:
\begin{align}
\label{edmn050}
\Pi^{Had}(p,q)&=\frac{\lambda^2_{\mathrm{P_{cc}}}}{[(p+q)^2-m^2_{\mathrm{P_{cc}}}][p^2-m^2_{\mathrm{P_{cc}}}]}
  \bigg[\Big(f_1(q^2)+f_2(q^2)\Big)\Big(
  2 (\varepsilon . p) \pslash -
  m_{\mathrm{P_{cc}}}\,\eslash \pslash
  -m_{\mathrm{P_{cc}}}\,\eslash \qslash
  +\eslash\pslash\qslash
  \Big)+ \cdots \bigg],\\
  \nonumber\\
\Pi^{Had}_{\mu\nu}(p,q)&=\frac{\lambda_{_{{\mathrm{P_{cc}^*}}}}^{2}}{[(p+q)^{2}-m_{_{{\mathrm{P_{cc}^*}}}}^{2}][p^{2}-m_{_{{\mathrm{P_{cc}^*}}}}^{2}]} 
\bigg[  g_{\mu\nu}\pslash\eslash\qslash \,F_{1}(q^2) 
-m_{{\mathrm{P_{cc}^*}}}g_{\mu\nu}\eslash\qslash\,F_{2}(q^2)
-
\frac{F_{3}(q^2)}{4m_{{\mathrm{P_{cc}^*}}}}q_{\mu}q_{\nu}\eslash\qslash \nonumber\\
&
-
\frac{F_{4}(q^2)}{4m_{{\mathrm{P_{cc}^*}}}^3}(\varepsilon.p)q_{\mu}q_{\nu}\pslash\qslash 
+
\cdots 
\bigg]. \label{final phenpart1}
\end{align}

To compute the $F_M(Q^2)$ and $G_M(Q^2)$ magnetic form factors, where the $F_M(q^2)$ and $G_M(q^2)$ are magnetic form factors for the $\rm{P_{cc}}$ and $\rm{P_{cc}^*}$ pentaquarks, respectively, for these doubly-charmed pentaquarks, it is necessary to express the form factors in terms of the previously defined $F_i(Q^2)$. These expressions are provided below:
\begin{align}
\label{edmn07}
F_M(q^2) &= f_1(q^2) + f_2(q^2),\\
G_{M}(q^2) &= \left[ F_1(q^2) + F_2(q^2)\right] ( 1+ \frac{4}{5}
\eta ) -\frac{2}{5} \left[ F_3(q^2)  \right]
+\left[
F_4(q^2)\right] \eta \left( 1 + \eta \right),
\end{align}  
where $\eta
= -\frac{q^2}{4m^2_{{\mathrm{P_{cc}^*}}}}$.  
The aforementioned expressions permit the determination of the electromagnetic form factors of these doubly-charmed pentaquarks. However, as we are working with a real photon ($q^2 =0$), we require expressions for these form factors in terms of the magnetic dipole moment. The following expressions provide the relevant details:
\begin{align}
\label{edmn08}
\mu_{\mathrm{P_{cc}}} &= \frac{ e}{2\, m_{\mathrm{P_{cc}}}} \,F_M( 0),
\\
\mu_{{\mathrm{P_{cc}^*}}}&=\frac{e}{2m_{{\mathrm{P_{cc}^*}}}}G_{M}(0),
\end{align}
with $F_{M}(0) = f_1(0)+f_2(0)$, and $G_{M}(0)= F_1(0)+F_2(0)$.

In the context of  QCD parameters, the quark field is first contracted with Wick's theorem, and then the operator product expansion (OPE) is performed. Following the contraction of the quark field operators on the QCD side, the correlation function is obtained regarding the propagators of both the light and heavy quarks, as well as the distribution amplitudes (DAs) of the photon. To illustrate, for the $\rm{D \Sigma_c}$ state with $I = 1/2$, the result of these manipulations is acquired in the following manner:
\begin{align}
\label{QCD1}
\Pi^{\rm{QCD}-\mathrm{ D \Sigma_c}}(p,q)&= \frac{i}{3}\varepsilon^{abc} \varepsilon^{a^{\prime}b^{\prime}c^{\prime}}\, \int d^4x \, e^{ip\cdot x} 
\nonumber\\
& 
 \langle 0\mid \Big\{ \mbox{Tr}\Big[\gamma_5  S_{c}^{dd^\prime}(x) \gamma_5 S_{d}^{d^\prime d}(-x) \Big]  
\mbox{Tr}\Big[\gamma_{\alpha} S_d^{bb^\prime}(x) \gamma_{\beta}  
  \widetilde S_{u}^{aa^\prime}(x)\Big]
(\gamma^{\alpha}\gamma_5 S_{c}^{cc^\prime}(x) \gamma_5  \gamma^{\beta})
 \nonumber\\
&     
  +2 \mbox{Tr}\Big[\gamma_5  S_{c}^{dd^\prime}(x) \gamma_5 S_{u}^{d^\prime d}(-x) \Big]  
\mbox{Tr}\Big[\gamma_{\alpha} S_u^{bb^\prime}(x) \gamma_{\beta}  
  \widetilde S_{u}^{aa^\prime}(x)\Big]
(\gamma^{\alpha}\gamma_5 S_{c}^{cc^\prime}(x) \gamma_5  \gamma^{\beta})
 \nonumber\\
& + 2 \mbox{Tr}\Big[\gamma_5  S_{c}^{dd^\prime}(x) \gamma_5 S_{u}^{d^\prime d}(-x) \Big]  
\mbox{Tr}\Big[\gamma_{\alpha} S_u^{ba^\prime}(x) \gamma_{\beta}  
  \widetilde S_{u}^{ab^\prime}(x)\Big]
(\gamma^{\alpha}\gamma_5 S_{c}^{cc^\prime}(x) \gamma_5  \gamma^{\beta})
 \Big\}
\mid 0 \rangle _\gamma \,,
\end{align}
where   
$\widetilde{S}_{Q(q)}^{ij}(x)=CS_{Q(q)}^{ij\mathrm{T}}(x)C$. The relevant light ($S_{q}(x)$) and heavy ($S_{Q}(x)$) quark propagators are expressed in the following manner~\cite{Yang:1993bp, Belyaev:1985wza}:
\begin{align}
\label{edmn13}
S_{q}(x)&= S_q^{free}(x) 
- \frac{\langle \bar qq \rangle }{12} \Big(1-i\frac{m_{q} \xslash}{4}   \Big)
- \frac{ \langle \bar qq \rangle }{192}
m_0^2 x^2  \Big(1 
  -i\frac{m_{q} \xslash}{6}   \Big)
+\frac {i g_s~G^{\mu \nu} (x)}{32 \pi^2 x^2} 
\bigg[\rlap/{x} 
\sigma_{\mu \nu} +  \sigma_{\mu \nu} \rlap/{x}
 \bigg],\\
%
S_{Q}(x)&=S_Q^{free}(x)
-\frac{m_{Q}\,g_{s}\, G^{\mu \nu}(x)}{32\pi ^{2}} \bigg[ (\sigma _{\mu \nu }{\xslash}
+{\xslash}\sigma _{\mu \nu }) 
    \frac{K_{1}\big( m_{Q}\sqrt{-x^{2}}\big) }{\sqrt{-x^{2}}}
 +2\sigma_{\mu \nu }K_{0}\big( m_{Q}\sqrt{-x^{2}}\big)\bigg],
 \label{edmn14}
\end{align}%
with  
\begin{align}
 S_q^{free}(x)&=\frac{1}{2 \pi x^2}\Big(i \frac{\xslash}{x^2}- \frac{m_q}{2}\Big),\\
 S_Q^{free}(x)&=\frac{m_{Q}^{2}}{4 \pi^{2}} \bigg[ \frac{K_{1}\big(m_{Q}\sqrt{-x^{2}}\big) }{\sqrt{-x^{2}}}
+i\frac{{\xslash}~K_{2}\big( m_{Q}\sqrt{-x^{2}}\big)}
{(\sqrt{-x^{2}})^{2}}\bigg],
\end{align}
where $m_0$ is defined through the quark-gluon mixed condensate $ m_0^2= \langle 0 \mid \bar  q\, g_s\, \sigma_{\mu\nu}\, G^{\mu\nu}\, q \mid 0 \rangle / \langle \bar qq \rangle $, $G^{\mu\nu}$ is the gluon field-strength tensor, and $K_n$'s being the modified second type Bessel functions.  Here, we use the following integral representation
 of the  modified second type Bessel function,     
\begin{equation}\label{b2}
K_n(m_Q\sqrt{-x^2})=\frac{\Gamma(n+ 1/2)~2^n}{m_Q^n \,\sqrt{\pi}}\int_0^\infty dt~\cos(m_Qt)\frac{(\sqrt{-x^2})^n}{(t^2-x^2)^{n+1/2}}.
\end{equation}

In the QCD representation, two distinct contributions to the correlation function can be identified. The first is the perturbative contribution, which arises when a photon is radiated at short distances. The second is the non-perturbative contribution, which occurs when a photon is radiated at long distances. To obtain contributions coming from short distances, it is necessary to use the following formula, 
\begin{align}
\label{free}
S^{free}(x) \rightarrow \int d^4y\, S^{free} (x-z)\,\rlap/{\!A}(z)\, S^{free} (z)\,,
\end{align}
and the four propagators that survive in Eq.~(\ref{QCD1}) are considered to be free. This amounts to taking $\bar T_4^{\gamma} (\underline{\alpha}) = 0$ and $S_{\gamma} (\underline {\alpha}) = \delta(\alpha_{\bar q})\delta(\alpha_{q})$ as the light-cone distribution amplitude in the three particle distribution amplitudes (see Ref. \cite{Li:2020rcg}). To obtain contributions coming from the long distances, it is recommended that the following approach  should be employed,
 \begin{align}
\label{edmn21}
S_{\alpha\beta}^{ab}(x) \rightarrow -\frac{1}{4} \Big[\bar{q}^a(x) \Gamma_i q^b(0)\Big]\Big(\Gamma_i\Big)_{\alpha\beta},
\end{align}
and the four propagators that remain in Eq.~(\ref{QCD1}) are considered to be full propagators. Here $\Gamma_i$ $=$ $\{\textbf{1}$, $\gamma_5$, $\gamma_\mu$, $i\gamma_5 \gamma_\mu$, $\sigma_{\mu\nu}/2\}$. When the procedures outlined in Eq. (\ref{edmn21}) are performed, new expressions, such as $\langle \gamma(q)\vel \bar{q}(x) \Gamma_i G_{\alpha\beta}q(0) \ver 0\rangle$ and $\langle \gamma(q)\vel \bar{q}(x) \Gamma_i q(0) \ver 0\rangle$, emerge as prerequisites for the continued analysis.  These terms, which are defined by photon wave functions, are of great importance for the evaluation of long-distance contributions (for details see Ref.~\cite{Ball:2002ps}). The computations of the QCD representation of the correlation functions are conducted using the expressions in Eqs.~(\ref{QCD1})-(\ref{edmn21}). These expressions are then transferred to the momentum space through the application of the Fourier transform. The QCD representation of the magnetic dipole moments is obtained through the application of the aforementioned, tedious and technical procedures.

One can then establish the QCD light-cone sum rules for the magnetic dipole moments, which are based on the quark-hadron duality, that the correlation functions obtained at the hadronic and quark-gluonic levels must be equal to each other. At this point, it is necessary to undertake one additional step. That is to increase the contribution of the ground state and at the same time reduce the effect of the contributions of the continuum and higher states.  To this end, we have applied the continuum subtraction and Borel transformation techniques following the standard procedure of the QCD light-cone sum rules approach. The outcomes yielded by the application of the aforementioned procedures for magnetic dipole moments are as follows:
\begin{align}
\label{edmn15}
&\mu_{ D \Sigma_c} \,\lambda^2_{ D \Sigma_c}\, m_{ D \Sigma_c}  =e^{\frac{m^2_{ D \Sigma_c}}{\rm{M^2}}}\, \rho_1 (\rm{M^2},\rm{s_0}),\\
&\mu_{ D \Sigma_c^*} \,\lambda^2_{ D \Sigma_c^*}\, m_{ D \Sigma_c^*}  =e^{\frac{m^2_{ D \Sigma_c^*}}{\rm{M^2}}}\, \rho_2 (\rm{M^2},\rm{s_0}),\\
&\mu_{ D^* \Sigma_c} \,\lambda^2_{ D^* \Sigma_c}\, m_{ D^* \Sigma_c} =e^{\frac{m^2_{ D^* \Sigma_c}}{\rm{M^2}}}\, \rho_3 (\rm{M^2},\rm{s_0}).
\end{align}
As the forms of the $\rho_i(\rm{M^2},\rm{s_0})$ functions are similar, for illustrative purposes, we only present the explicit form of the $\rho_1(\rm{M^2},\rm{s_0})$ function in the following manner: 
\begin{align}
\label{sumrules}
 \rho_1 (\rm{M^2},\rm{s_0}) &= \frac{m_c}{2 ^{26}\times 3^2 \times 5^3\times 7^2 \pi^7} \Big[ 3 (e_d + 9 e_u) \big(149 I[0, 7] + 427 I[1, 6]\big) + 
 5 e_c \big(239 I[0, 7] + 1337 I[1, 6]\big)\Big] \nonumber\\
 &-\frac {m_c^2\langle g_s^2 G^2\rangle \langle \bar q q \rangle f_ {3\gamma } } {2^{20} \times 3^3 \pi^3} (e_d + 9 e_u) I_2[\psi^a] I[0, 2]\nonumber\\
   & - \frac {m_c^2 \langle \bar q q \rangle m_ 0^2 f_ {3 \gamma  }  } {2^{19}\times 3^4 \times 5 \pi^3}(e_d + 9 e_u)  I_2[\psi^a] I[0,3]  \nonumber\\
 & -\frac {m_c\, \langle \bar q q \rangle^2   } {2^{18} \times 3^2 \times 5^2 \pi^3}  \Big[ (e_d + 9 e_u)\Big ( 
      15  I[0, 4] h_\gamma[u_ 0] + 
       2 \chi I[0, 5] \varphi_\gamma[u_ 0]\Big)\Big]\nonumber\\
       & + \frac {m_c\,\langle g_s^2 G^2\rangle f_ {3 \gamma} } {2^{27}\times 3^5 \times 5 \pi^5} (e_d + 9 e_u)\Big [ 
   63 I_ 1[\mathcal V] + 80 (-9 I_2[\psi^a] + 4 \psi^a[u_ 0])\Big] I[
   0, 4]\nonumber\\
   & + \frac { m_c^2 \langle \bar q q \rangle f_ {3 \gamma} } {2^{19}\times 3^4 \times 5 \pi^3} (e_d + 9 e_u)\Big[ 9 \big(I_ 1[\mathcal V] - 4 I_2[\psi^a]\big) I[0, 
      4] + 22 I[0, 4] \psi^a[u_ 0] \Big]\nonumber\\
      & + \frac { m_c \,
   f_ {3\gamma } } {2 ^{24}\times 3^2 \times 5^2 \times 7^2 \pi^5} (e_d + 
     9 e_u)\Big [ \big(5 I_ 1[\mathcal V] + 10 I_2[\psi^a] + 
      7 \psi^a[u_ 0]\big) I[0, 6]\Big],
\end{align}
where the $I[n,m]$ and $I_i[\mathcal{F}]$ functions are given  as:
\begin{align}
 I[n,m]&= \int_{\mathcal M}^{\rm{s_0}} ds \, s^n (s-\mathcal M)^m\,e^{-s/\rm{M^2}},~
 \nonumber\\
 I_1[\mathcal{F}]&=\int D_{\alpha_i} \int_0^1 dv~ \mathcal{F}(\alpha_{\bar q},\alpha_q,\alpha_g) \delta'(\alpha_ q +\bar v \alpha_g-u_0),\nonumber\\
   I_2[\mathcal{F}]&=\int_0^1 du~ \mathcal{F}(u)\delta'(u-u_0),
 \end{align}
 with $\mathcal M = 4\,m_c^2$,  and $\mathcal{F}$ being the relevant photon DAs.

\end{widetext}

\section{Numerical analysis of the magnetic dipole moments}\label{numerical}

In this section, we will perform numerical analyses of the expressions related to the magnetic dipole moments for doubly-charmed pentaquarks. To do so, we require a set of input parameters, which are presented in Table \ref{inputparameter}.   From sum rules in Eq. (\ref{sumrules}), it follows that the photon DAs are also required. Their explicit expressions are borrowed from Ref.~\cite{Ball:2002ps}.
 %
 \begin{table}[htb!]
	\addtolength{\tabcolsep}{10pt}
	\caption{Input parameters in this work~\cite{
	Ioffe:2005ym,Workman:2022ynf,Rohrwild:2007yt,Wang:2024brl,Narison:2018nbv}
	.}
	\label{inputparameter}
\begin{tabular}{l|c|ccccc}
               \hline\hline
Inputs& Values&Unit\\
                                        \hline\hline
$m_c$&$ 1.27 \pm 0.02$&GeV  
                        \\
$m_{ D \Sigma_c}$&$  4.30^{+0.07}_{-0.08}$&GeV
                        \\
$m_{ D \Sigma_c^*}$&$ 4.38^{+0.07}_{-0.07}$&GeV
                        \\
$m_{ D^* \Sigma_c} $&$  4.46^{+0.08}_{-0.08}$&GeV
                       \\
$m_0^{2} $&$ 0.8 \pm 0.1 $&\,\,GeV$^2$  
                       \\
$\chi $&$ -2.85 \pm 0.5 $&\,\,\,\,\,GeV$^{-2}$ 
                       \\
$\langle \bar qq\rangle $&$ (-0.24 \pm 0.01)^3 $&\,\,GeV$^3$
                       \\
$ \langle g_s^2G^2\rangle  $&$ 0.48 \pm 0.14 $&\,\,GeV$^4$ 
                       \\
$\lambda_{ D \Sigma_c}  $&$ (2.46^{+0.34}_{-0.32})\times 10^{-3} $&\,\,GeV$^6$
                       \\
$ \lambda_{ D \Sigma_c^*}   $&$ (1.61^{+0.21}_{-0.20})\times 10^{-3} $&\,\,GeV$^6$ 
                       \\
$\lambda_{ D^* \Sigma_c}   $&$(3.13^{+0.41}_{-0.39})\times 10^{-3} $&\,\,GeV$^6$
                       \\
                                      \hline\hline
 \end{tabular}
\end{table}

\begin{figure}[htb!]
\subfloat[]{\includegraphics[width=0.45\textwidth]{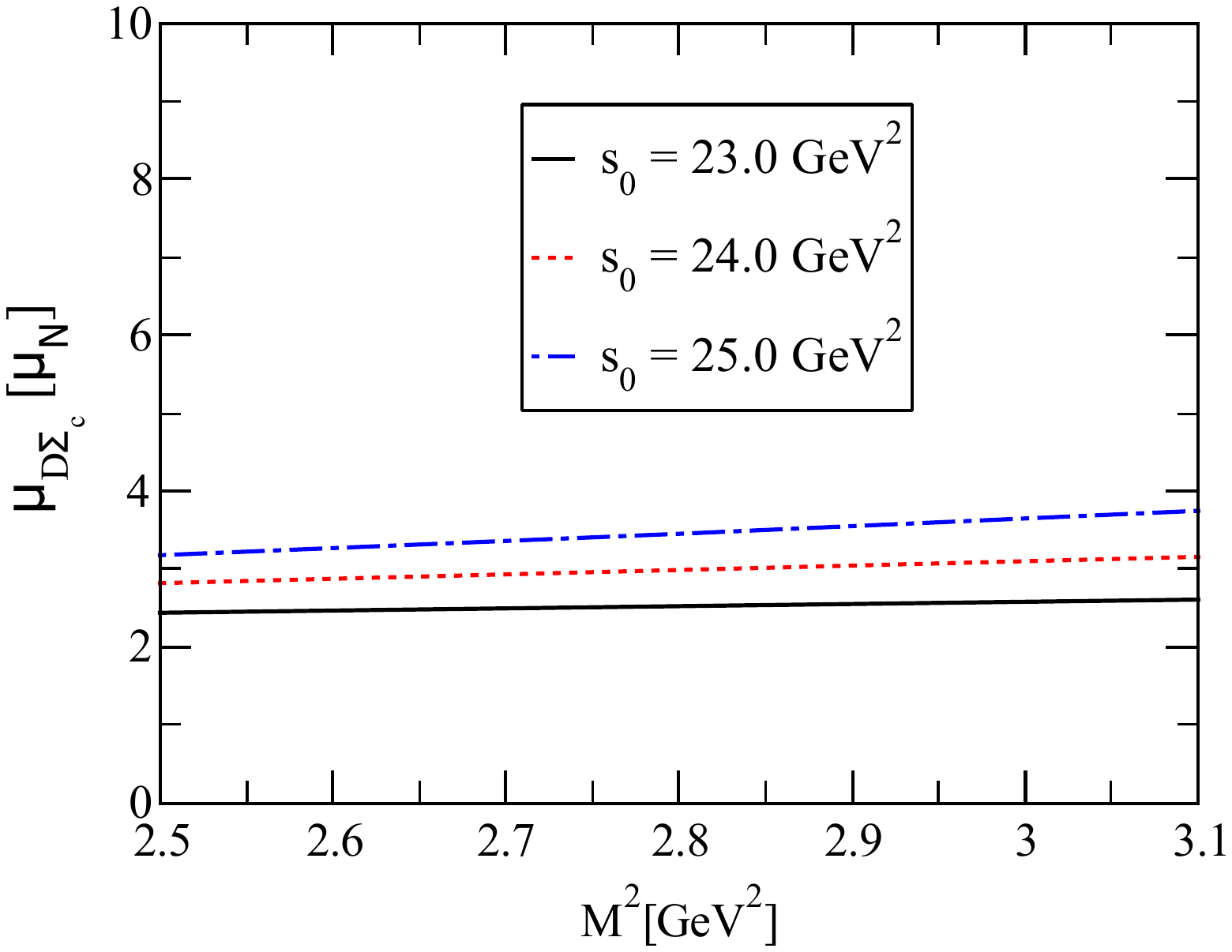}}~~~~~~~~~~~~
\subfloat[]{\includegraphics[width=0.45\textwidth]{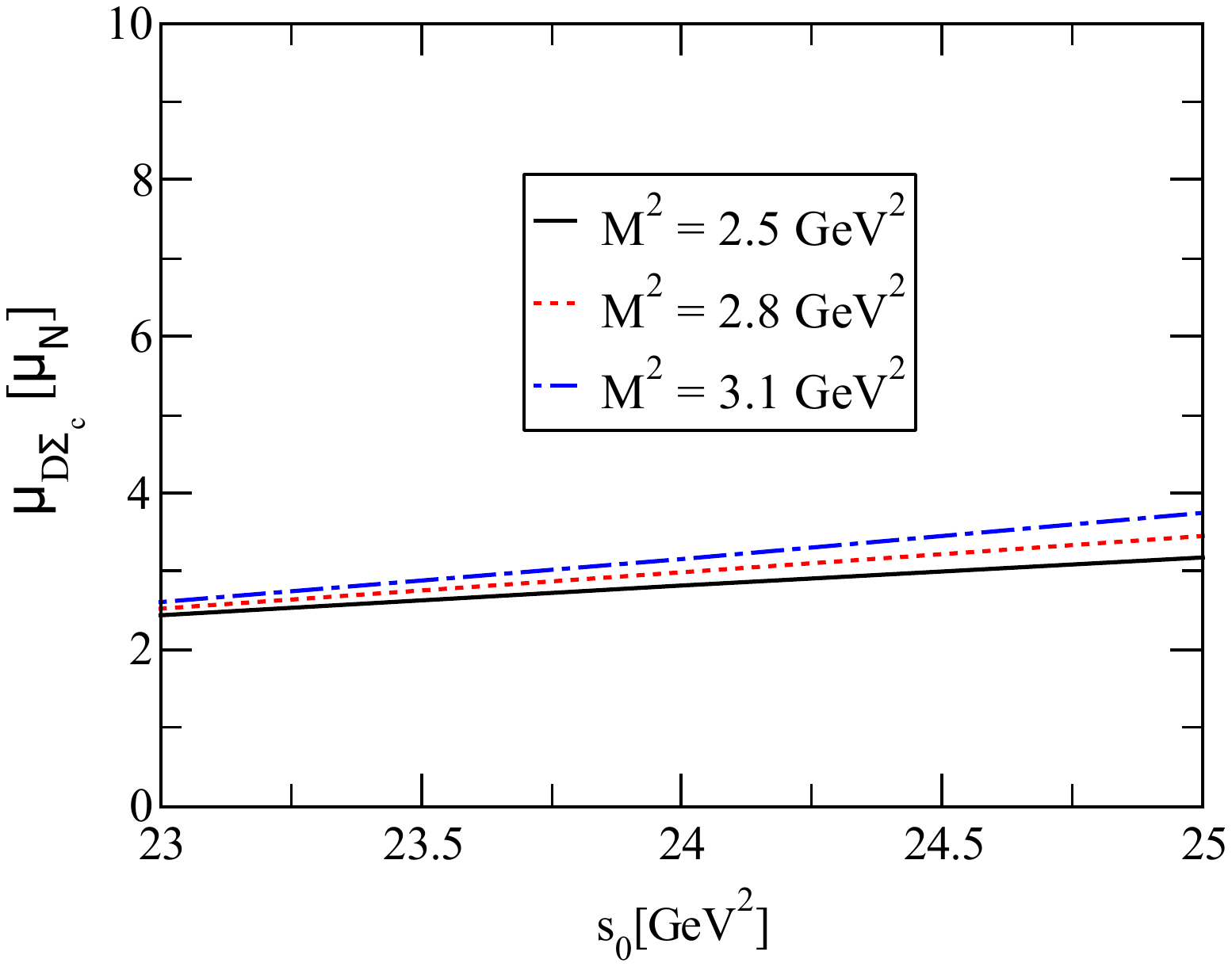}}\\
\subfloat[]{\includegraphics[width=0.45\textwidth]{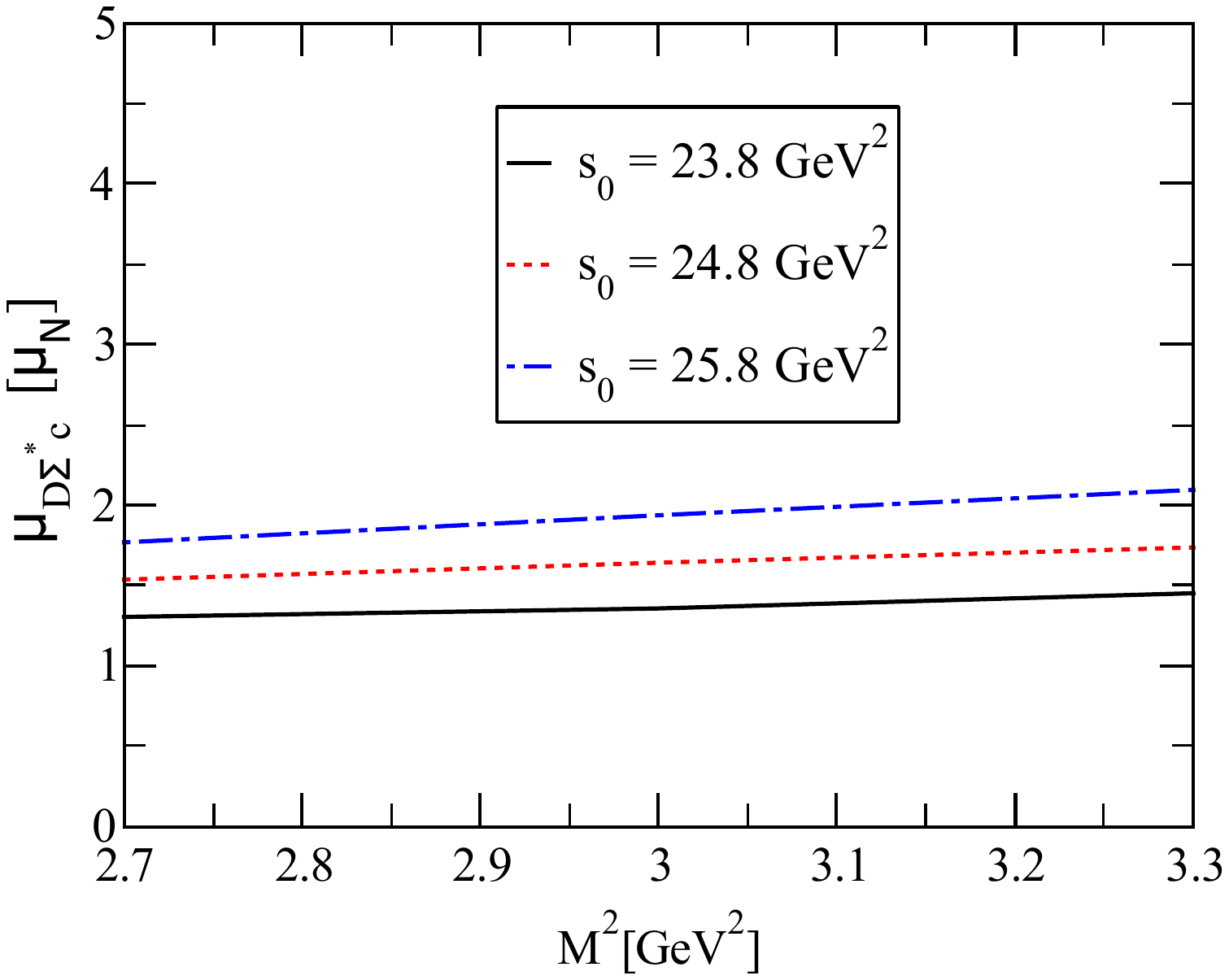}}~~~~~~~~~~~~
\subfloat[]{\includegraphics[width=0.45\textwidth]{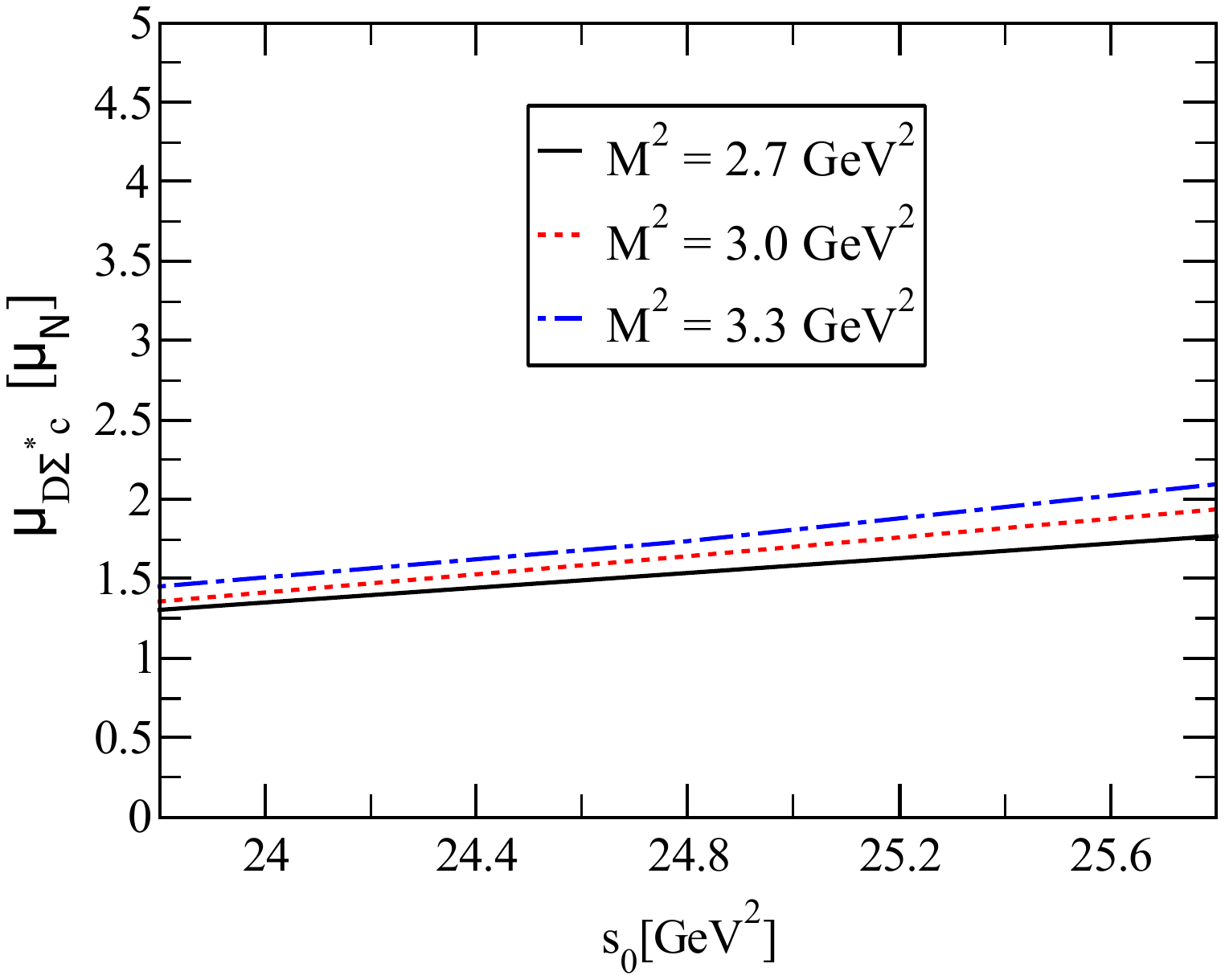}}\\
\subfloat[]{\includegraphics[width=0.45\textwidth]{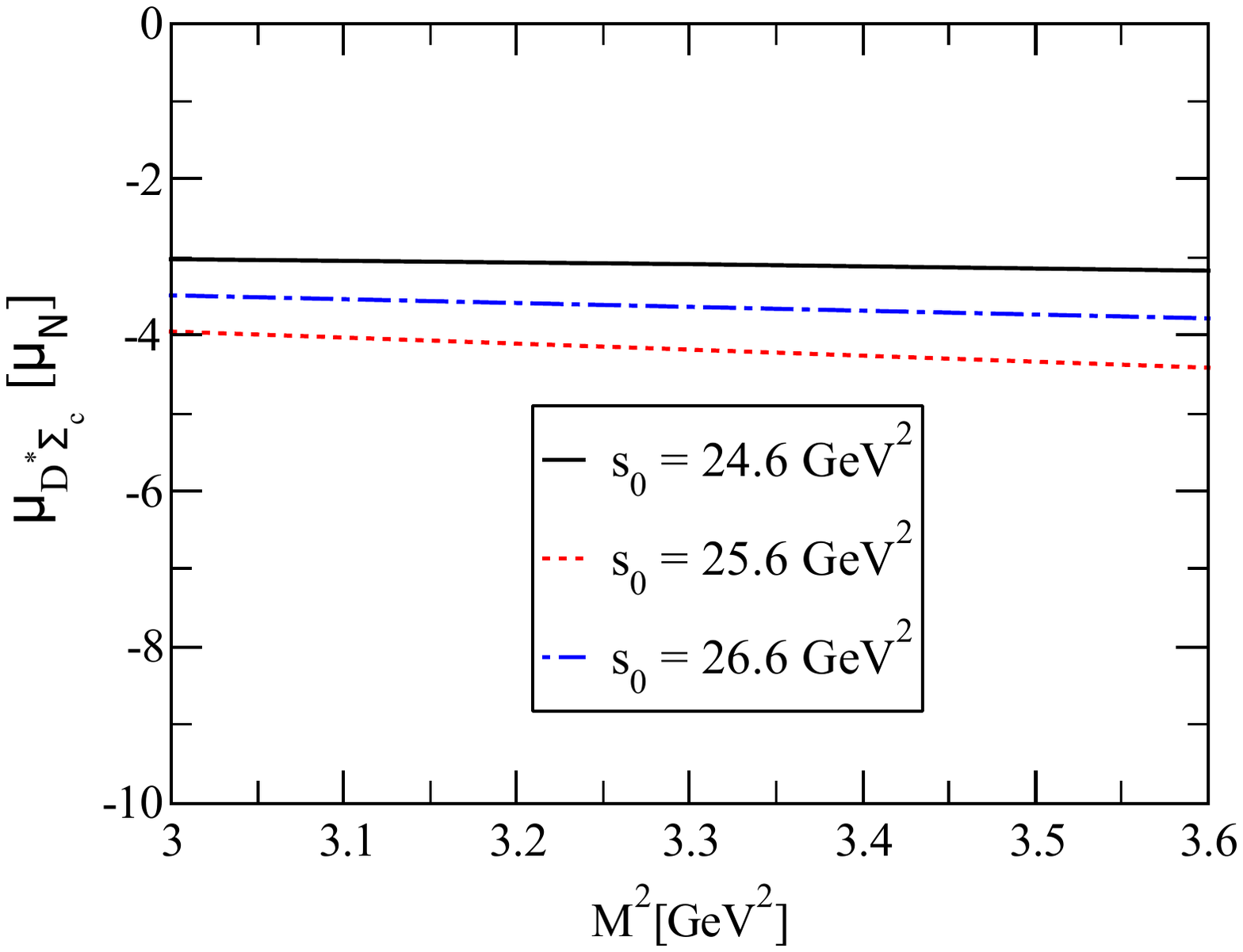}}~~~~~~~~~~~~
\subfloat[]{\includegraphics[width=0.45\textwidth]{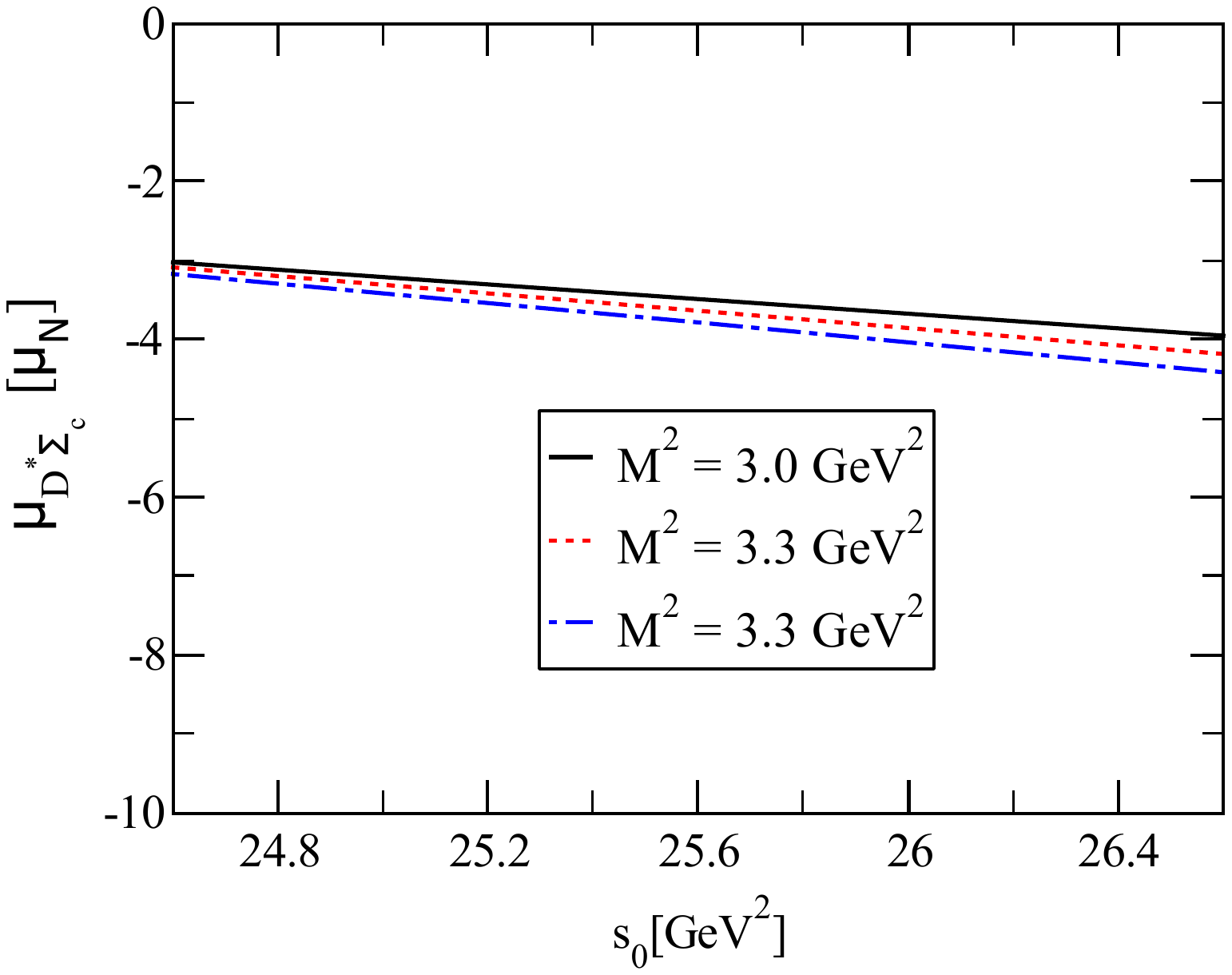}}
 \caption{The magnetic dipole moments of the doubly-charmed pentaquarks versus $\rm{M^2}$ (left panel), and $\rm{s_0}$ (right panel).}
 \label{Msqfig1}
  \end{figure}

In addition to the aforementioned parameters, there are two further parameters to consider: the Borel parameter, denoted by $\rm{M^2}$, and the threshold parameter, denoted by $\rm{s_0}$. The results are analyzed under the standard constraints of the QCD sum rule method. 
These constraints are known as pole dominance (PC) and convergence of OPE (CVG) are expressed in the following manner:
\begin{align}
 \mbox{PC} &=\frac{\rho_i (\rm{M^2},\rm{s_0})}{\rho_i (\rm{M^2},\infty)},~~~~~~~~
 \mbox{CVG} =\frac{\rho_i^{\mbox{DimN}} (\rm{M^2},\rm{s_0})}{\rho_i (\rm{M^2},\rm{s_0})},
 \end{align}
 where $\rho_i^{\mbox{DimN}} (\rm{M^2},\rm{s_0})$ is  the highest dimensional term in the OPE of $\rho_i (\rm{M^2},\rm{s_0})$. 
 Following the aforementioned requirements, we have established the working windows for the auxiliary parameters $\rm{M^2}$ and $\rm{s_0}$. These intervals are presented in Table \ref{parameter} for the states under study. Additionally, the values of PC and CVG obtained from the analyses for each state are included in the aforementioned table. 
For the sake of completeness, Fig. \ref{Msqfig1} depicts the dependence of the magnetic dipole moment of the doubly-charmed pentaquarks on the Borel mass parameter, $\rm{M^2}$, and $\rm{s_0}$. As can be observed from this figure, the variation of magnetic dipole moments concerning $\rm{M^2}$ is relatively stable. Despite the high variation compared to $\rm{s_0}$, this variation remains within the error limits of the method employed.
%
\begin{table}[htb!]
	\addtolength{\tabcolsep}{10pt}
	\caption{Working regions of $\rm{s_0}$ and  $\rm{M^2}$ together with the PC and CVG for the magnetic dipole moments of the doubly-charmed pentaquarks.}
	\label{parameter}
	\begin{ruledtabular}
\begin{tabular}{l|cccccc}
               \\
Pentaquarks & $\rm{s_0}$ (GeV$^2$) & $\rm{M^2}$ (GeV$^2$)&  ~~  PC ($\%$) ~~ & ~~  CVG  
 ($\%$) \\
 \\
                                        \hline\hline
                                        \\
$ D \Sigma_c$ & $23.0-25.0$ & $2.5-3.1$ & $37.66$ &  $<2.0$  
                        \\
                        \\
$ D \Sigma^*$& $23.8-25.8$ & $2.7-3.3$ & $34.52$ &  $<2.5$  
                       \\
                       \\
$ D^* \Sigma_c$ & $24.6-26.6$ & $3.0-3.6$ & $33.08$ &  $<2.5$   
                      \\
                       \\
 \end{tabular}
\end{ruledtabular}
\end{table}
%

After the numerical analyses, we collect the predicted magnetic moments for $\rm{J^P} =\frac{1}{2}^-$ and $\frac{3}{2}^-$ doubly-charmed pentaquarks are presented in Table \ref{sonuc}. For completeness, the electric quadrupole and magnetic octupole moments of the $\rm{J^P} = \frac{3}{2}^-$ doubly-charmed pentaquarks are also included.  The errors in the results can be attributed to the inherent uncertainties associated with the $\rm{s_0}$ and  $\rm{M^2}$, in addition to the various input parameters enumerated in Table \ref{inputparameter}, along with the input parameters employed in the photon DAs.  
%
%
  \begin{table}[htp]
	\addtolength{\tabcolsep}{10pt}
	\caption{Numerical values of the magnetic dipole moments for the doubly-charmed pentaquarks with isospin-1/2. As a by-product, the electric quadrupole ($\mathcal Q$) and the magnetic octupole ($\mathcal O$) moments of the relevant pentaquarks are also given.}
	\label{sonuc}
		\begin{ruledtabular}
\begin{tabular}{l|cccccc}
                \\
Pentaquarks &$\mu\,(\mu_N)$& $\mathcal Q$\,(fm$^2$)($\times 10^{-2}$) & $\mathcal O$\,(fm$^3$)($\times 10^{-3}$)  \\
 \\
                                        \hline\hline
                                        \\
$ D \Sigma_c$ & $2.98^{+0.76}_{-0.54}$ & - & -  
                        \\
                        \\
$ D \Sigma_c^*$ & $1.65^{+0.45}_{-0.34}$ &$-2.09^{+0.36}_{-0.33}$  &  $-0.09^{+0.02}_{-0.02}$
                       \\
                        \\
$ D^* \Sigma_c$ & $-3.63^{+0.79}_{-0.60}$ & $-0.79^{+0.15}_{-0.13}$ & $1.07^{+0.11}_{-0.09}$    
                      \\
                       \\
 \end{tabular}
\end{ruledtabular}
\end{table}
%

Several key points can be summarized from the numerical results presented in Table \ref{sonuc}:
\begin{itemize}
  
\item The initial observation from the analysis is that the $D^{(*)0} \Sigma_c^{(*)++}$ component of the interpolating current provides the primary contribution. This contribution is approximately $90\%$ for the $D \Sigma_c$ and  $D \Sigma_c^*$ pentaquarks, while in the case of $ D^* \Sigma_c$ it is approximately $80\%$.

\item  When we analyze the individual quark contributions (this can be done with the help of the charge factors $e_u$, $e_d$, and $e_c$), we find that: In the $D \Sigma_c$ case, the magnetic moment is governed by light quarks (terms proportional to $e_u$ ($90\%$) and $e_d$ ($10\%$)). In contrast, in the $D \Sigma_c^*$ case, the opposite is true, with the c-quark (terms proportional to $e_c$ ($88\%$)) governing the analysis. Finally, in the case of the $D^* \Sigma_c$, both light and heavy quarks contribute almost equally to the results, with the former contributing terms proportional to $e_c$ ($53\%$), secondly the contributing terms proportional to $e_u$ ($40\%$), and the least contributing terms proportional to $e_d$ ($7\%$).

\item The results of the magnetic dipole moment sign obtained for the  $D \Sigma_c$ and   $D \Sigma_c^*$ pentaquarks are positive, while the sign of the $ D^* \Sigma_c$ pentaquark is negative. The main reason for this may be that different diquark components of the interpolating currents used in the calculations produce different results in the analysis.  Upon a detailed examination of the analysis, it has been determined that the terms contributing the most to the $D \Sigma_c$ and   $D \Sigma_c^*$ states are proportional to those with positive signs, whereas, in the  $D^* \Sigma_c$ state, the terms contributing the most are proportional to those with negative signs. This discrepancy in the signs of these pentaquarks has been observed. 

\item The magnitude of magnetic dipole moments may be employed to provide insights into the experimental accessibility of such entities. The magnetic dipole moments obtained for the doubly-charmed molecular pentaquarks are notably large, a consequence of the double electric charge. Their magnitude suggests that these results may be accessible in future experiments. 

\item The results obtained for the higher multipole moments are found to be non-zero, which suggests that the charge distributions of the $ D \Sigma_c^{*}$ and $ D^{*} \Sigma_c$ pentaquarks are non-spherical. The electric quadrupole moment sign for the $D \Sigma_c^*$ and $ D^* \Sigma_c$ pentaquarks is predicted to be negative, indicating that the geometric shape of these pentaquarks is oblate.

\end{itemize}

To ensure the comprehensiveness of the analysis, it is necessary to compare the results obtained with those presented in the existing literature. 
In Ref.~\cite{Zhou:2022gra}, the magnetic dipole moments of the $D \Sigma_c$,  $D \Sigma_c^*$, and $ D^* \Sigma_c$ pentaquarks have been calculated through the constituent quark model in the molecular configuration. The magnetic dipole moments are obtained as  $1.737\mu_N$, $3.163\mu_N$, and $1.178\mu_N$ for the $D \Sigma_c$,  $D \Sigma_c^*$, and $ D^* \Sigma_c$ pentaquarks, respectively. A comparison of the quark model results with those obtained in this study reveals significant discrepancies, indicating that the results are incompatible with each other. The discrepancy in the results may be attributed to the use of different models and interactions, as well as the consideration of different sets of parameters. To get a more conclusive picture of these results, further studies are encouraged.

  \begin{table}[htp]
	\addtolength{\tabcolsep}{10pt}
	\caption{Numerical values of the electromagnetic  multipole moments for the doubly-charmed pentaquarks with isospin-3/2.}
	\label{sonuc2}
		\begin{ruledtabular}
\begin{tabular}{l|cccccc}
                \\
Pentaquarks &$\mu\,(\mu_N)$& $\mathcal Q$\,(fm$^2$)($\times 10^{-2}$) & $\mathcal O$\,(fm$^3$)($\times 10^{-3}$)  \\
 \\
                                        \hline\hline
                                        \\
$ D \Sigma_c$ & $3.78^{+0.94}_{-0.70}$ & - & -  
                        \\
                        \\
$ D \Sigma_c^*$ & $2.08^{+0.57}_{-0.43}$ &$-2.64^{+0.46}_{-0.40}$  &  $-0.12^{+0.03}_{-0.03}$
                       \\
                        \\
$ D^* \Sigma_c$ & $-4.59^{+0.99}_{-0.76}$ & $-1.00^{+0.18}_{-0.17}$ & $1.35^{+0.14}_{-0.12}$    
                      \\
                       \\
 \end{tabular}
\end{ruledtabular}
\end{table}
%
%

As a final step of our analyses in this study, we have also presented the results of the magnetic dipole and higher multipole moments of the isospin-3/2 partners of these double-charmed pentaquarks. The results are listed in Table \ref{sonuc2}. It should be noted that the masses of both isospin-1/2 and isopin-3/2 doubly-charmed pentaquarks have been obtained to be identical~\cite{Wang:2024brl}. Nevertheless, the results of the magnetic dipole and higher multipole moment calculations differ by approximately $27\%$ to $33\%$. As known well, the magnetic dipole moments of the hadrons usually may reflect their inner structures. Therefore, the magnetic dipole and higher multipole moment results presented in this study may be used to distinguish between isospin-1/2 and isospin-3/2 doubly-charmed pentaquarks.

\section{summary and concluding notes}\label{summary}
  
To gain insight into the nature of the controversial and as yet incompletely understood exotic states, we are conducting a comprehensive investigation into their electromagnetic properties. The magnetic dipole moment of a hadron represents a fundamental aspect of the dynamical properties of the state and contains significant information regarding its deeper underlying structure. 
Motivated by the discoveries of the hidden-charm pentaquarks, the doubly-charmed $T_{cc}^+(3875)$ state, the magnetic dipole moments of the isospin eigenstate of $ D \Sigma_c$, $ D \Sigma_c^{*}$ and $ D^{*} \Sigma_c$ doubly-charmed pentaquarks are extracted, which are directly related to the inner organization of the relevant states. The magnetic dipole moments of these states have been evaluated through the QCD light-cone sum rules technique with isospin spin-parity $\rm{I(J^P)} = \frac{1}{2}(\frac{1}{2}^-)$, $\rm{I(J^P)} = \frac{1}{2}(\frac{3}{2}^-)$ and $\rm{I(J^P)} = \frac{1}{2}(\frac{3}{2}^-)$,  for $ D \Sigma_c$, $ D \Sigma_c^{*}$ and $ D^{*} \Sigma_c$ doubly-charmed pentaquarks respectively. The magnetic dipole moments obtained for the doubly-charmed molecular pentaquarks are notably large, a consequence of the double electric charge. Their magnitude suggests that these results may be accessible in future experiments. Furthermore, we have also extracted the electric quadrupole and the magnetic octupole moments of the $ D \Sigma_c^{*}$ and $ D^{*} \Sigma_c$ doubly-charmed pentaquarks. The results obtained for the higher multipole moments are found to be non-zero, indicating that the charge distributions of the $ D \Sigma_c^{*}$ and $ D^{*} \Sigma_c$ pentaquarks are non-spherical. As it is known, the electric quadrupole moment allows for the acquisition of information regarding the geometric shape of the system under study. The electric quadrupole moment sign for the $D \Sigma_c^*$ and $ D^* \Sigma_c$ pentaquarks is predicted to be negative, indicating that the geometric shape of these pentaquarks is oblate.

The electromagnetic characteristics of the pentaquarks with doubly-charm can be employed to ascertain substantial knowledge regarding the size and geometric shape of the hadrons. To interpret the properties of hadrons concerning quark-gluon degrees of freedom, it is essential to determine the physical observables in question. As a topic with significant potential for research, the electromagnetic properties of unconventional states warrant further attention from both experimental and theoretical researchers, which can enrich our understanding of the true nature of hadronic unconventional states.  This can also deepen our understanding of the complex and elusive non-perturbative character of the strong interaction in the low-energy regime.
 
%
%


\bibliographystyle{elsarticle-num}
\bibliography{Dsigmac_pentaquarksMM.bib}

\end{document}